\documentclass[usenatbib]{mn2e}

\usepackage{amssymb,amsmath}
\usepackage{graphicx}
\usepackage{multirow}
\usepackage{natbib}
\def\citeN{\citet}
\def\cite{\citep}

\def\b50cm{50}

\usepackage{url}
\usepackage{float}
\usepackage{bm}
\footnotesize
\newdimen\minuswidth    
\setbox0=\hbox{$-$}
\minuswidth=\wd0
\catcode`@=\active
\def@{\kern\minuswidth}
\newdimen\digitwidth    
\setbox0=\hbox{\rm0}
\digitwidth=\wd0
\catcode`!=\active
\def!{\kern\digitwidth}
\normalsize

\usepackage[usenames,dvipsnames]{color}

\footnotesize
\newdimen\digitwidth    
\setbox0=\hbox{\rm0}
\digitwidth=\wd0
\catcode`!=\active
\def!{\kern\digitwidth}
\normalsize
\title[Correction of DM variations in pulsar timing]{Measurement and correction of variations in interstellar dispersion in high-precision pulsar timing}
\makeatletter
\author[M.~J.~Keith et al.]
{M.~J.~Keith$^{1}$\thanks{Email: mkeith@pulsarastronomy.net},
W.~Coles$^2$,
R.~M.~Shannon$^1$,
G.~B.~Hobbs$^1$,
R.~N.~Manchester$^1$, \newauthor
M.~Bailes$^3$,
N.~D.~R.~Bhat$^{3,4}$,
S.~Burke-Spolaor$^{5}$,
D.~J.~Champion$^6$,
A.~Chaudhary$^1$, \newauthor
A.~W.~Hotan$^1$,
J.~Khoo$^1$,
J.~Kocz$^{3,7}$,
S.~Os{\l}owski$^{3,1}$,
V.~Ravi$^{8,1}$,
J.~E.~Reynolds$^1$, \newauthor
J.~Sarkissian$^1$,
W.~van Straten$^3$,
D.~R.~B.~Yardley$^{9,1}$
\\
$^1$ Australia Telescope National Facility, CSIRO Astronomy \& Space Science, P.O. Box 76, Epping, NSW 1710, Australia\\
$^2$ Electrical and Computer Engineering, University of California at San Diego, La Jolla, CA, U.S.A. \\
$^3$ Centre for Astrophysics and Supercomputing, Swinburne University of Technology, P.O. Box 218, Hawthorn, VIC 3122, Australia \\
$^4$ International Centre for Radio Astronomy Research, Curtin University, Bentley, WA 6102, Australia \\
$^5$ NASA Jet Propulsion Laboratory, California Institute of Technology, 4800 Oak Grove Drive, Pasadena, CA 91109, USA \\
$^6$ Max-Planck-Institut f{\"u}r Radioastronomie, Auf dem H{\"u}gel 69, 53121, Bonn, Germany \\
$^7$ Harvard-Smithsonian Centre for Astrophysics, 60 Garden Street, Cambridge, MA 02138, U.S.A.\\
$^8$ School of Physics, University of Melbourne, Vic 3010, Australia \\
$^9$ Sydney Institute for Astronomy, School of Physics A29, The University of Sydney, NSW 2006, Australia\\
}
\makeatother

\let\leqslant=\leq 
\date{}
\begin{document}

\maketitle
\newcommand{\setthebls}{
}

\setthebls

\begin{abstract}
Signals from radio pulsars show a wavelength-dependent delay due to dispersion in the interstellar plasma.
At a typical observing wavelength, this delay can vary by tens of microseconds on five-year time scales, far in excess of signals of interest to pulsar timing arrays, such as that induced by a gravitational-wave background.
Measurement of these delay variations is not only crucial for the detection of such signals, but also provides an unparallelled measurement of the turbulent interstellar plasma at au scales.

In this paper we demonstrate that without consideration of wavelength-independent red-noise, `simple' algorithms to correct for interstellar dispersion can attenuate signals of interest to pulsar timing arrays.
We present a robust method for this correction, which we validate through simulations, and apply it to observations from the Parkes Pulsar Timing Array.
Correction for dispersion variations comes at a cost of increased band-limited white noise. We discuss scheduling to minimise this additional noise, and factors, such as scintillation, that can exacerbate the problem.

Comparison with scintillation measurements confirms previous results that the spectral exponent of electron density variations in the interstellar medium often appears steeper than expected.
We also find a discrete change in dispersion measure of PSR J1603$-$7202 of $\sim 2 \times 10^{-3}$ cm$^{-3}$pc for about 250 days. We speculate that this has a similar origin to the `extreme scattering events' seen in other sources.
In addition, we find that four pulsars show a wavelength-dependent annual variation, indicating a persistent gradient of electron density on an au spatial scale, which has not been reported previously.

\end{abstract}


\section{Introduction}
The fundamental datum of a pulsar timing experiment is the time of arrival (ToA) of a pulse at an observatory.
In practise, the ToA is referred to the solar-system barycentre in a standard time frame (e.g., barycentric coordinate time).
This barycentric arrival time can be predicted using a `timing model' for the pulsar.
The difference between the barycentric ToAs and the arrival times predicted by the timing model are termed residuals.
The timing model can be refined using a least-squares fitting procedure to minimise the residuals, as performed by, e.g., the {\sc Tempo2} software \cite{hem06}.
Since the timing model is always incomplete at some level, we always see some level of post-fit residuals, which are typically a combination of `white' noise due to the uncertainty in the ToA measurement and `red' (i.e., time-correlated) signal.
For the majority of known pulsars the dominant red signal is caused by the intrinsic instability of the pulsar, and termed `timing noise' (e.g., \citealp{hlk10}). However, the subset of millisecond pulsars are stable enough that other red signals are potentially measurable \cite{vbc+09}.
Pulsar timing array projects, such as the Parkes Pulsar Timing Array (PPTA; \citealp{mhb+12}), aim to use millisecond pulsars to detect red signals such as: errors in the atomic time standard \cite{hcm+12}; errors in the Solar System ephemeris \cite{chm+10}; or the effect of gravitational waves \cite{yhj+10,ych+11,vlj+11}.
Each of these signals can be distinguished by the spatial correlation, i.e., how pulsars in different directions on the sky are affected.
However, at typical observing wavelengths and time-spans, the variation of the dispersive delay due to turbulence in the ionised interstellar medium (ISM) dominates such signals \cite{yhc+07}.
Fortunately for pulsar timing experiments, these delays can be measured and corrected using observations at multiple wavelengths.

The dispersive group delay is given by
\begin{equation}
\label{dm_eq}
t_{\rm DM} = \lambda^2 \left [ \frac{e^2}{2\pi m_ec^3} \int_{path} \!\!\!\!\!\! n_e (l) \mbox{d}l \right ],
\end{equation}
where $\lambda$ is the barycentric radio wavelength\footnote{To avoid confusion, in this paper we will use wavelength for the radio wavelength and frequency to describe the fluctuation of time variable processes.}.
The path integral of electron density is the time-variable quantity.
In pulsar experiments this is termed `dispersion measure', DM, and given in units of cm$^{-3}$pc.
In principle, the instantaneous DM can be computed from the difference of two arrival times from simultaneous observations at different wavelengths, or more generally by fitting to any number of observations at more than one wavelength.

The question of estimation and correction of DM$(t)$ has previously been considered by \citeN{yhc+07}. They chose a `best' pair of wavelengths from those available and estimated the DM at every group of observations. These observation groups were selected by hand, as was the choice of wavelengths. Regardless of how the analysis is done, the estimated DM always contains white noise from differencing two observations, and correcting the group delay always adds that white noise to the arrival times. However the DM$(t)$ variations are red, so they only need to be corrected at frequencies below the `corner frequency' at which the power spectrum of the DM-caused fluctuations in group delay is equal to the power spectrum of the white noise in the DM$(t)$ estimate. To minimise the additional white noise, they smoothed the DM$(t)$ estimates over a time $T_{\rm s}$ to create a low-pass filter which cuts off the DM variations, and the associated white noise, at frequencies above the corner frequency. In this way, they avoided adding white noise at high frequencies where the DM-correction was unnecessary. Of course the added `white' noise is no longer white; it is white below the corner frequency, but zero above the corner frequency. 

Here we update this algorithm in two ways. We use all the observed wavelengths to estimate DM$(t)$ and we integrate the smoothing into the estimation algorithm automatically. Thus, the algorithm can easily be put in a data `pipeline'. We show the results of applying this new algorithm to the PPTA data set, which is now about twice as long as when it was analysed by \citeN{yhc+07}.
Additionally, we demonstrate that our algorithm is unbiased in the presence of wavelength-independent red signals, e.g., from timing noise, clock error, or gravitational waves; and we show that failure to include wavelength-independent red signals in the estimation algorithm will significantly reduce their estimated amplitude.

\section{Theory of Dispersion Removal}

We assume that an observed timing residual is given by $t_{{\rm OBS}} = t_{\rm CM} + t_{\rm DM} (\lambda / \lambda_{\rm REF} )^2$ where $t_{\rm CM}$ is the common-mode, i.e., wavelength-independent delay and $t_{\rm DM}$ is the dispersive delay at some reference wavelength $\lambda_{\rm REF}$. Then with observations at two wavelengths we can solve for both $t_{\rm CM}$ and $t_{\rm DM}$.
\begin{equation}
\label{eq1}
\tilde{t}_{\rm DM} = (t_{{\rm OBS},1} - t_{{\rm OBS},2}) \lambda_{\rm REF}^2 /(\lambda_1^2 - \lambda_2^2),
\end{equation}
\begin{equation}
\tilde{t}_{\rm CM} = (t_{{\rm OBS},2} \lambda_1^2 - t_{{\rm OBS},1} \lambda_2^2 ) /(\lambda_1^2 - \lambda_2^2).
\end{equation}
In a pulsar timing array, $t_{\rm CM}$ would represent a signal of interest, such as a clock error, an ephemeris error, or the effect of a gravitational wave. The dispersive component $t_{\rm DM}$ would be of interest as a measure of the turbulence in the ISM, but is a noise component for other purposes. It is important to note that $\tilde{t}_{\rm DM}$ is independent of $t_{\rm CM}$ so one can estimate and correct for the effects of dispersion regardless of any common-mode signal present.
In particular, common-mode red signals do not cause any error in $\tilde{t}_{\rm DM}$.

If more than two wavelengths are observed, solving for $t_{\rm CM}$ and $t_{\rm DM}$ becomes a weighted least-squares problem, and the standard deviation of the independent white noise on each observation is needed to determine the weighting factors. For wavelength $i$, we will denote the white noise by $t_{W,i}$ and its standard deviation by $\sigma_i$ so the observed timing residual is modelled as
\begin{equation}
\label{eq4}
t_{{\rm OBS},i} = t_{\rm CM} + t_{\rm DM} (\lambda_i / \lambda_{\rm REF} )^2 + t_{W,i}.
\end{equation}
The weighted least-squares solutions, which are minimum variance unbiased estimators, are
\begin{eqnarray}
\label{e5}
\tilde{t}_{\rm DM} &=& \lambda_{\rm REF}^2  \Big( \sum_i 1/ \sigma_i^2  \sum_i t_{{\rm OBS},i} \lambda_i^2 / \sigma_i^2 - \nonumber \\ 
 & & \sum_i \lambda_i^2 / \sigma_i^2  \sum_i t_{{\rm OBS},i} / \sigma_i^2 \Big) / \Delta \\
\label{e6}
\tilde{t}_{\rm CM} &=& \Big( \sum_i \lambda_i^4 / \sigma_i^2  \sum_i t_{{\rm OBS},i} / \sigma_i^2 - \nonumber \\
& & \sum_i \lambda_i^2 / \sigma_i^2  \sum_i t_{{\rm OBS},i} \lambda_i^2 / \sigma_i^2 \Big) / \Delta. 
\end{eqnarray}
Here $\Delta$ is the determinant of the system of equations,
\begin{equation}
\Delta = {\sum_i 1/ \sigma_i^2  \sum_i \lambda_i^4 / \sigma_i^2 -  \Bigl( \sum_i \lambda_i^2 / \sigma_i^2 \Bigr)^2}. \nonumber \\
\end{equation}
If one were to model only the dispersive term $t_{\rm DM}$, the weighted least-squares solution would become
\begin{equation}
\tilde{t}_{\rm DM}= \lambda_{\rm REF}^2 \frac{ \sum_i t_{{\rm OBS},i} \lambda_i^2 / \sigma_i^2 } 
{\sum_i \lambda_i^4 / \sigma_i^2 }.
\end{equation}
However if a common-mode signal is present, this solution is biased. The expected value is
\begin{equation}
\langle \tilde{t}_{\rm DM} \rangle = t_{\rm DM} + t_{\rm CM} \lambda_{\rm REF}^2 \frac{ \sum_i  \lambda_i^2 / \sigma_i^2 } 
{\sum_i \lambda_i^4 / \sigma_i^2 }.
\end{equation}
Some of the `signal' $t_{\rm CM}$ is absorbed into $\tilde{t}_{\rm DM}$ reducing the effective signal-to-noise ratio and degrading the estimate of DM. We will demonstrate this bias using simulations in Section \ref{sec:simulations}.

It is important to note that the dispersion estimation and correction process is linear - the estimators $\tilde{t}_{\rm DM}$ and $\tilde{t}_{\rm CM}$ are linear combinations of the residuals. The corrected residuals $t_{{\rm OBS},{\rm cor},i} = t_{{\rm OBS},i} - (\lambda_i /\lambda_{\rm REF} )^2 \tilde{t}_{\rm DM}$, are also linear combinations of the residuals. We can easily compute the white noise in any of these quantities from the white noise in the residuals.
For example, we can collect terms in Equations (\ref{e5}) and (\ref{e6}) obtaining $\tilde{t}_{\rm DM} = \sum_i a_i t_{{\rm OBS},i}$ and $\tilde{t}_{\rm CM} = \sum_i b_i t_{{\rm OBS},i}$, where
\begin{eqnarray}
\label{e9}
a_i &=& \lambda_{\rm REF}^2  \Big(  \lambda_i^2 / \sigma_i^2  \sum_j 1/\sigma^2_j  -   1 / \sigma_i^2 \sum_j \lambda_j^2 / \sigma_j^2  \Big) / \Delta \\
\label{e10}
b_i &=&   \Big(   1/\sigma_i^2 \sum_j \lambda_j^4 / \sigma_j^2  -  \lambda_i^2 / \sigma_i^2  \sum_j \lambda_j^2 / \sigma_j^2 \Big)  / \Delta.
\end{eqnarray}
Then, the white noise variances of the estimators can be written as $\sigma_{\rm TDM}^2 = \sum_i a_i^2 \sigma_i^2$ and $\sigma_{\rm TCM}^2 = \sum_i b_i^2 \sigma_i^2$.

The actual PPTA observations are not simultaneous at all frequencies, so we cannot normally apply Equations (\ref{e5}) and (\ref{e6}) directly \cite{mhb+12}.
We discuss how the least squares solutions for $\tilde{t}_{\rm DM}$ and $\tilde{t}_{\rm CM}$ can be obtained by including them in the timing model in the next section.
However it is useful to have an analytical estimate of the power spectral density of the white noise that one can expect in these estimators and in the corrected residuals.
At each wavelength $\lambda_i$ we have a series of $N_i$ error estimates $\sigma_{ij}$.
The variance of the weighted mean is $\sigma_{mi}^2 = 1/\sum_j {1/\sigma_{ij}^2}$.
This is the same as if we had a different number $N$ of observations at this wavelength each of variance $\sigma^2 = \sigma_{mi}^2 N$.
Thus, for planning purposes we can compute $\sigma_{mi}$ for each wavelength and conceptually resample each wavelength with an
arbitrary number ($N$) of samples.
Equations (\ref{e5}), (\ref{e6}), (\ref{e9}), and (\ref{e10}) are invariant under scaling of all $\sigma_i$ by the same factor so one can obtain the coefficients $a_i$ and $b_i$ using $\sigma_{mi}$ in place of $\sigma_i$ so the actual number ($N_i$) of samples need not enter the equations.

If one had a series of $N$ samples over a time span of $T_{{\rm OBS}}$ each with variance $\sigma^2$, the spectral density of the white noise would be $P_w = 2 T_{{\rm OBS}}\ \sigma^2 /N = 2 T_{{\rm OBS}} \ \sigma_m^2$. We can extend this to a weighted white noise spectral density using the variance of the weighted mean. 
So the power spectral densities $P_{w,i}$ play the same role as $\sigma_i^2$ in Equations (\ref{e5}), (\ref{e6}), (\ref{e9}) and (\ref{e10}).
The coefficients \{$a_i$\} and \{$b_i$\} are functions of $\lambda_i$ and $P_{w,i}$. Then we find $P_{w,{\rm TDM}} = \sum_i a_i^2 P_{w,i}$ and $P_{w,{\rm TCM}} = \sum_i b_i^2 P_{w,i}$. 

Perhaps the most important property of these estimators is that $P_{w,{\rm TCM}}$ is less than or equal to the white noise spectrum of the corrected residuals $P_{w,{\rm cor},i}$ in any band. Equality occurs when there are only two wavelengths.
The values of $P_{w,i}$, $P_{w,{\rm cor},i}$, $P_{w,{\rm TDM}}$ and $P_{w,{\rm TCM}}$ are given for the PPTA pulsars in Table \ref{psd_table}. Here $P_{w,{\rm TDM}}$ is given at the reference wavelength of $20\,$cm.

The situation is further complicated by red noise which depends on wavelength, but not as $\lambda^2$. For example, diffractive angular scattering causes variations in the group delay, which scale as the scattered pulse width, i.e. approximately as $\lambda^{4}$ \cite{ric77}. Clearly such noise will enter the DM correction process. It can have the unfortunate effect that scattering variations, which are stronger at long wavelengths, enter the short wavelength corrected residuals even though they are negligible in the original short wavelength data. This will be discussed in more detail in Section \ref{longer_wavelength}.

\begin{table*}
\caption{\label{psd_table}
The estimated power spectral density before ($P_w$) and after ($P_{w,{\rm cor}}$) correction of the white noise for each PPTA pulsar at each of the three wavelengths, and the expected white noise power spectral density in the `common mode' signal ($P_{w,\rm{TCM}}$) and in $t_{\rm DM}$ at 20 cm ($P_{w,\rm{TDM}}$), all expressed relative to the power spectral density of the uncorrected 20-cm residuals.
Also shown is the effect of optimising the observing time, expressed as the ratio of $P_{w,\rm{TCM}}$ estimated for optimal observing and $P_{w,\rm{TCM}}$ with the current observing strategy ($\alpha=0.5$), and $\alpha_{\rm opt}$ the optimal fraction of time spent using the dual 10- and \b50cm-cm observing system.
}
\begin{tabular}{lcccccccccc}
\hline \hline

Source &    $P_{w,20}$  & \multirow{2}{*}{$\frac{P_{w,10}}{P_{w,20}}$}   &  \multirow{2}{*}{$\frac{P_{w,50}}{P_{w,20}}$} 
& \multirow{2}{*}{$\frac{P_{w,{\rm cor},10}}{P_{w,20}}$}  & \multirow{2}{*}{$\frac{P_{w,{\rm cor},20}}{P_{w,20}}$}   &  \multirow{2}{*}{$\frac{P_{w,{\rm cor},50}}{P_{w,20}}$} & \multirow{2}{*}{$\frac{P_{w,\rm{TCM}}}{P_{w,20}}$} & \multirow{2}{*}{$\frac{P_{w,\rm{TDM}}}{P_{w,20}}$} & \multirow{2}{*}{$\frac{P_{w,\rm{TCM}}(\alpha_{\rm opt})}{P_{w,\rm{TCM}}(\alpha_{0.5})}$} & \multirow{2}{*}{$\alpha_{\rm opt}$}\\
       & (yr$^3$) &\\
\hline	    
J0437$-$4715 &  $1.4\!\times\!10^{-31}$ & 1 & 19 & 1.3 & 1.4 & 7.2 & 1 & 1.1             & 0.6 & 1.0\\
J0613$-$0200 &  $6.2\!\times\!10^{-29}$ & 18 & 0.68 & 18 & 1.8 & 1.7 & 1.7 & 0.17        & 0.7 & 0.2\\
J0711$-$6830 &  $2.3\!\times\!10^{-28}$ & 4.8 & 6.5 & 5.1 & 2 & 2.2 & 1.8 & 0.69         & 1.0 & 0.5\\
J1022$+$1001 &  $5.3\!\times\!10^{-28}$ & 2.3 & 2.5 & 2.4 & 1.5 & 1.3 & 1.2 & 0.28       & 1.0 & 0.6\\
J1024$-$0719 &  $9.2\!\times\!10^{-29}$ & 33 & 12 & 33 & 2.9 & 3.2 & 2.9 & 1.4           & 1.0 & 0.5\\
J1045$-$4509 &  $4.8\!\times\!10^{-28}$ & 17 & 3.1 & 17 & 2 & 2 & 1.9 & 0.43             & 0.9 & 0.3\\
J1600$-$3053 &  $2.7\!\times\!10^{-29}$ & 2.9 & 15 & 3.3 & 2.1 & 4.6 & 1.8 & 1.2         & 0.9 & 0.8\\
J1603$-$7202 &  $4.3\!\times\!10^{-28}$ & 6.3 & 1.5 & 6.5 & 1.7 & 1.5 & 1.5 & 0.24       & 0.9 & 0.3\\
J1643$-$1224 &  $9.2\!\times\!10^{-29}$ & 4.3 & 2.8 & 4.6 & 1.7 & 1.6 & 1.5 & 0.35       & 1.0 & 0.4\\
J1713$+$0747 &  $6.7\!\times\!10^{-30}$ & 2.1 & 33 & 2.6 & 2.2 & 15 & 1.9 & 1.9          & 0.6 & 1.0\\
J1730$-$2304 &  $4.2\!\times\!10^{-28}$ & 1.2 & 2.1 & 1.4 & 1.4 & 0.94 & 0.84 & 0.22     & 0.8 & 1.0\\
J1732$-$5049 &  $5.1\!\times\!10^{-28}$ & 25 & 10 & 26 & 2.7 & 3 & 2.7 & 1.2             & 1.0 & 0.5\\
J1744$-$1134 &  $2.8\!\times\!10^{-29}$ & 3.9 & 12 & 4.4 & 2.2 & 3.7 & 1.9 & 1.1         & 0.9 & 0.7\\
J1824$-$2452A &  $6.0\!\times\!10^{-29}$ & 30 & 40 & 32 & 5.5 & 8.7 & 5.4 & 4.1          & 0.9 & 0.7\\
J1857$+$0943 &  $8.4\!\times\!10^{-29}$ & 8.5 & 20 & 9.2 & 3.1 & 5.5 & 3 & 1.9           & 0.9 & 0.6\\
J1909$-$3744 &  $4.3\!\times\!10^{-30}$ & 0.51 & 6 & 0.61 & 1.1 & 1.5 & 0.5 & 0.42       & 0.6 & 1.0\\
J1939$+$2134 &  $4.2\!\times\!10^{-30}$ & 14 & 5.2 & 14 & 2.2 & 2.2 & 2.1 & 0.64         & 1.0 & 0.4\\
J2124$-$3358 &  $3.5\!\times\!10^{-28}$ & 19 & 2.8 & 19 & 2 & 1.9 & 1.9 & 0.4            & 0.9 & 0.3\\
J2129$-$5721 &  $1.1\!\times\!10^{-28}$ & 680 & 2.5 & 680 & 2.1 & 2.1 & 2.1 & 0.39       & 0.8 & 0.3\\
J2145$-$0750 &  $6.8\!\times\!10^{-29}$ & 8.2 & 15 & 8.8 & 2.7 & 4 & 2.6 & 1.5           & 1.0 & 0.6\\

\hline
\end{tabular}
\end{table*}

\section{Dispersion Correction Technique}
\label{technique}
Rather than solving for $t_{\rm CM}$ and $t_{\rm DM}$ for every group of observations, or re-sampling observations at each wavelength to a common rate, it is more practical to include parametrised functions for $t_{\rm CM} (t)$ and DM$(t)$ in the timing model used to obtain the timing residuals.
To provide a simple and direct parametrisation we use piece-wise linear models defined by 
fixed samples $t_{\rm CM} (t_j)$ and DM($t_j$) for j = 1,...,$N_s$.

It is also required to introduce some constraints into the least-squares fitting to prevent covariance with other model parameters. For example, the values of DM($t_j$) are naturally covariant with the mean dispersion measure parameter, ${\rm DM}_0$, which is central to the timing model. To eliminate this covariance, we implement the linear equality constraint that $\sum_{i=1} {\rm DM}(t_j) = 0$.
Additionally, the series $t_{\rm CM} (t_j)$ is covariant with the entire timing model, however in practise the sampling interval is such that it responds very little to any orbital parameters (in the case of binary systems). We constrain $t_{\rm CM}(t_j)$ to have no response to a quadratic polynomial, or to position, proper motion, and parallax.
These constraints are implemented as part of the least-squares fit in {\sc Tempo2}, as described in Appendix \ref{appendix2}


The choice of sampling interval, $T_{\rm s}$ is essentially the same as in \citeN{yhc+07}. The process of fitting to a piece-wise linear function is equivalent to smoothing the DM$(t)$ time series with a triangle function of base $2T_{\rm s}$. This is a low pass filter with  transfer function $H_{\rm tri}(f) = ( \sin (\pi f T_{\rm s})/\pi f T_{\rm s})^2$. We adjust $T_{\rm s}$ such that the pass band approximately corresponds to the corner frequency $f_c$ at which the power spectrum of the DM delays, $P_{\rm TDM}$, exceeds that of the white noise, $P_{w,{\rm TDM}}$. Note that this corner frequency is independent of reference wavelength at which $t_{\rm DM}$ is defined.

To determine this corner frequency we need an estimate of the power spectrum of $t_{\rm DM}$, so the process is inherently iterative.
We can obtain a first estimate of $P_{\rm TDM} (f)$ from the diffractive time scale, $\tau_{\rm diff}$, at the reference wavelength.
For signals in the regime of strong scattering, which includes all PPTA observations, $\tau_{\rm diff}$ is the time scale of the diffractive intensity scintillations.
For the PPTA pulsars, $\tau_{\rm diff}$ is usually of the order of minutes and can be estimated from a dynamic spectrum taken during normal observations (see e.g. \citealp{cwd+90}).

Rather than directly compute $P_{\rm TDM}$, it is attractive to begin with the structure function, which is a more convenient statistic for turbulent scattering processes and is more stable when only a short duration is available.
The structure function of $t_{\rm DM}$ is given by
\begin{equation}
D_{\rm TDM} (\tau) = \langle (t_{\rm DM} (t) - t_{\rm DM} (t + \tau ))^2 \rangle = (\lambda/2 \pi c )^2 D_\phi (\tau),
\end{equation}
where $D_\phi (\tau)$ is the phase structure function.
If we assume that the electron density power spectrum has an exponent of -11/3, i.e., Kolmogorov turbulence, then $D_\phi (\tau) = (\tau/\tau_{\rm diff} )^{5/3}$ \cite{fc90}.
The structure function $D_{\rm TDM} (\tau)$ can therefore be estimated from $\tau_{\rm diff}$, or directly from the $t_{\rm DM} (t)$ once known.

As described in Appendix \ref{appendix1} we can use the structure function at any time lag $\tau$ to obtain a model power spectrum using
\begin{equation}
\label{eq:pd}
P_{\rm TDM} (f) \simeq 0.0112 \, D_{\rm TDM} (\tau) \tau^{-5/3} {\rm (spy)}^{-1/3} f^{-8/3}
\end{equation}
The term (spy) is the number of seconds per year. Here $D_{\rm TDM}$ is in s$^2$, $\tau$ in s, $f$ is in yr$^{-1}$ and $P_{\rm TDM}$ is in yr$^3$.

The spectrum of the white noise can be estimated from the ToA measurement uncertainties as discussed in section 2.
However, often there are contributions to the white noise that are not reflected in the measurement uncertainties and so we prefer to estimate $P_w$ directly from the power spectrum of the residuals.

\section{Test on Simulated Observations}
\label{sec:simulations}
When dealing with real data sets it is not trivial to show that the DM-corrected residuals are `improved' over simply taking residuals from the best wavelength \cite{yhc+07}.
This is because much of the variations in DM are absorbed into the fit for the pulsar period and period derivative.
Therefore the root-mean-square deviation (RMS) of the residuals from a single wavelength may not decrease significantly even though the RMS of the DM$(t)$ variations that were removed is large.
To demonstrate that the proposed procedure can estimate and remove the dispersion, and that it is necessary to include the common-mode in the process, we perform two sets of simulations.

The observing parameters, i.e., $T_{\rm obs}$, $N_i$, $\sigma_{ij}$, $D_{\rm DM}(\tau)$, of both simulations are based on the observations of PSR J1909$-$3744 in the PPTA `DR1' data set \cite{mhb+12}. We find it useful to demonstrate the performance of the DM correction process in the frequency domain, but it is difficult to estimate power spectra of red processes if they are irregularly sampled. Therefore we first use simulations of regularly sampled observations with observing parameters similar to those of PSR J1909$-$3744 to demonstrate the performance of the DM correction algorithm. Then we will simulate the actual irregularly sampled observations of PSR J1909$-$3744 to show that the ultimate performance of the algorithm is the same as in the regularly sampled case.

\subsection{Regular sampling, equal errors}
We will compare the power spectra produced after fitting for DM$(t)$ with and without simultaneously fitting for a common-mode signal.
To generate the simulated data sets, we first generate idealised ToAs that have zero residual from the given timing model.
Then we add zero-mean stochastic perturbations to the ideal ToAs to simulate the three components of the model: (1) independent white noise, corresponding to measurement error; (2) wavelength independent red noise, corresponding to the common-mode; (3) wavelength dependent red noise representing DM$(t)$.

We simulate the measurement uncertainty with a white Gaussian process, chosen to match the high frequency power spectral density of the observed residuals. The simulated $P_w$ is $2.2 \times 10^{-30}, 4.3 \times 10^{-30}$ and $2.6 \times 10^{-29}\,{\rm yr}^3$ at 10, 20 and \b50cm$\,$cm respectively. For the common mode we choose a Gaussian process with a spectrum chosen to match a common model of the incoherent gravitational wave background (GWB), i.e. $P_{\rm GWB}(f) = (A_{\rm GWB}^2/12\pi^2) f^{-13/3}$ (\citealp{jhv+06,hjl+09}). For the DM we use a Gaussian process with a power spectrum $P_{\rm DM}(f) = A_{\rm DM} f^{-8/3}$, where $A_{\rm DM}$ is chosen to match the observed DM fluctuations in PSR J1909$-$3744 shown in Figures \ref{ppta_dm} and \ref{ppta_sf}, and the spectral exponent is chosen to match that expected for Kolmogorov turbulence \cite{fc90}. The levels of $P_{\rm TDM}$ and $P_{\rm GWB}$ are similar so that the same sample intervals can be used for both DM$(t_i )$ and $t_{\rm CM}(t_i )$, but this is not necessary in general and will not always be desirable.

\begin{figure*}
\includegraphics[width=14cm]{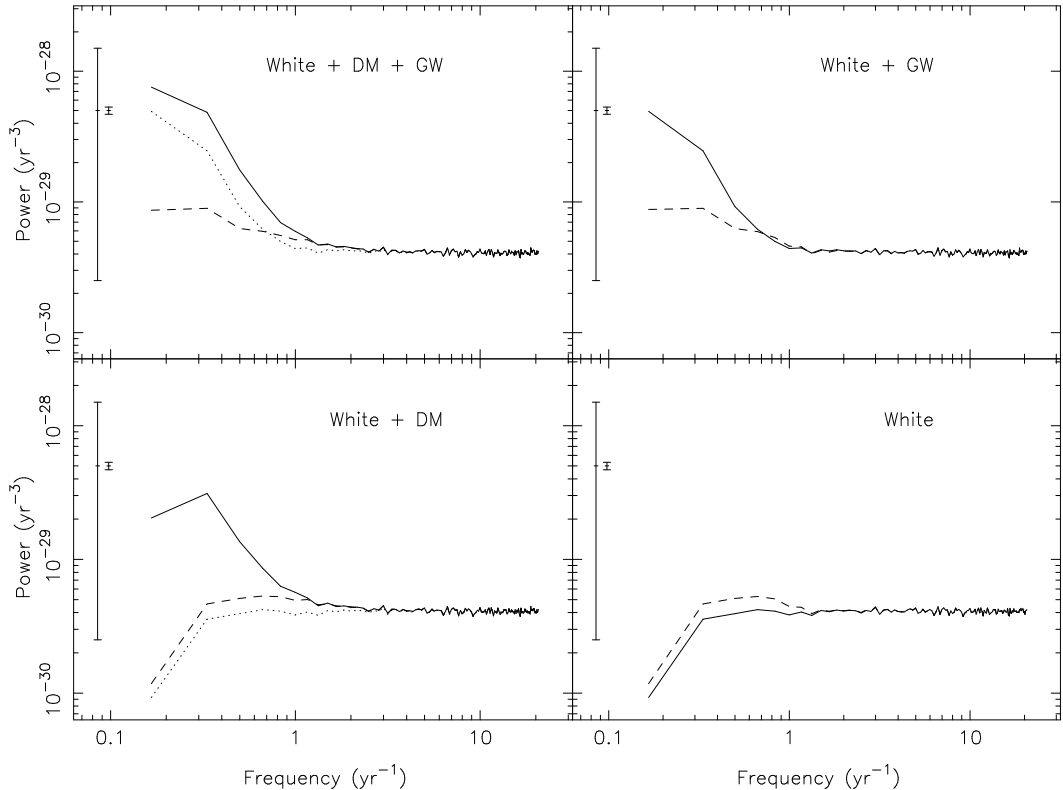}
\caption{
\label{50d_u_ncm}
Average power spectra of pre- and post-correction timing residuals, in the 20-cm band, with four combinations of signals. The solid line shows the pre-correction spectrum and dashed line shows the post-correction spectrum. For the cases where variations in DM are included in the simulation, the pre-correction spectrum without DM variations is shown with a dotted line.
Here the fitting routine uses the DM$(t)$ interpolated fitting routine, without fitting a common-mode signal.
The vertical bars on the left of each panel show the $90\%$ spectral estimation uncertainty for a single realisation (left-most bar) and the average of 1000 realisations (right bar).
}
\end{figure*}

\begin{figure*}
\includegraphics[width=14cm]{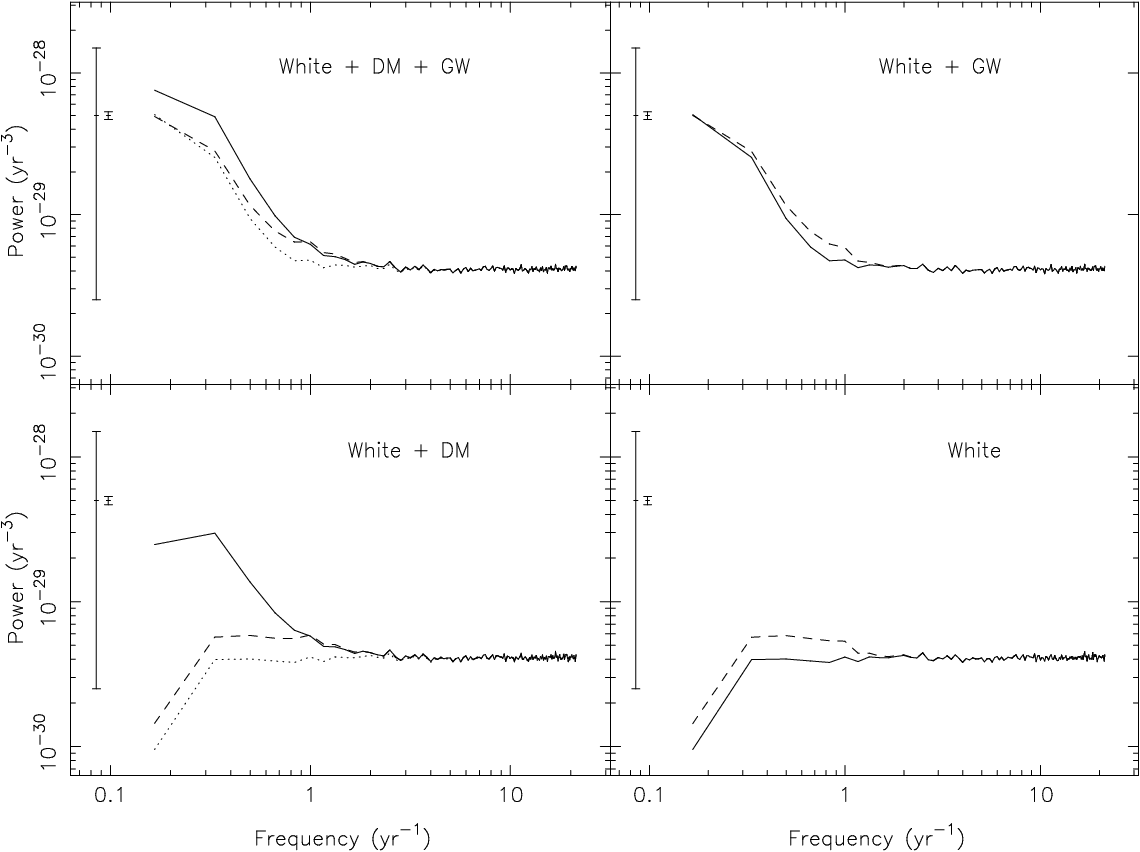}
\caption{
\label{50d_u}
As for Figure \ref{50d_u_ncm}, except the fitting routine uses the DM$(t)$ interpolated fitting routine in addition to the wavelength independent signal, C($t$).
}
\end{figure*}

For both algorithms, we estimate the pre- and post-correction power spectra of the 20-cm residuals in four noise regimes: $P_w$; $P_w + P_{\rm DM}$; $P_w + P_{\rm GWB}$; and $P_w + P_{\rm DM} + P_{\rm GWB}$. In order to minimise the statistical estimation error, we average together 1000 independent realisations of the spectra for each algorithm.
We note that although the averaged power spectra suggest that the input red noise signals are large, the noise on a single realisation is such that the red signals are at the limit of detection. To illustrate this, the 90\% confidence limits for both the 1000 spectrum average and for a single realisation, are shown on the power spectra in Figures \ref{50d_u_ncm} and \ref{50d_u}.

We show the effect of using the interpolated model for DM$(t)$, but not fitting for the common-mode signal $t_{\rm CM} (t)$, in Figure \ref{50d_u_ncm}. 
This algorithm is well behaved when the GWB is not present, as shown in the two lower panels. In this case the DM correction algorithm removes the effect of the DM variations if they are present and increases the white noise below the corner frequency by the expected amount. 
Importantly, when the model GWB is included, i.e., in the two top panels, a significant amount of the low-frequency GWB spectrum is absorbed into the DM correction. This is independent of whether or not DM variations are actually present because the DM correction process is linear.

We show the full algorithm developed for this paper, using interpolated models for both DM$(t)$ and the common-mode signal $t_{\rm CM} (t)$, in Figure \ref{50d_u}. 
One can see that the algorithm removes the DM if it is present, regardless of whether the GWB is present. It does not remove any part of the GWB spectrum. When the GWB is not present, as shown in the two lower panels, the algorithm remains well behaved.
As expected, it increases the white noise below the corner frequency by a larger factor than in the previous case. This is the `cost' of not absorbing some of the GWB signal into the DM correction.
Although it has a higher variance than for the previous case, our DM$(t)$ is the lowest variance unbiased estimator of the DM variations in the presence of wavelength-independent red noise.
This increase in white noise is unavoidable if we are to retain the signal from a GWB, or indeed any of the signals of interest in PTAs.

The power spectra presented in Figure \ref{50d_u} demonstrate that the algorithm is working as expected, in particular that it does not remove power from any wavelength-independent signals present in the data. We note, however, two limitations in these simulations: the regular sampling and equal errors are not typical of observations, nor have we shown that the wavelength-independent signal in the post-fit data is correlated with the input signal (since our power spectrum technique discards any phase information).
These limitations will be addressed in the next section.

\subsection{Irregular sampling, Variable error bars}
In order to test the algorithm in the case of realistic sampling and error bars, we repeated the simulations using the actual sampling and error bars for pulsar J1909$-$3744 from the PPTA. We use the same simulated spectral levels for the GW and DM as in the previous section. The results are also an average of 1000 realisations.

\begin{figure*}
\includegraphics[width=14cm]{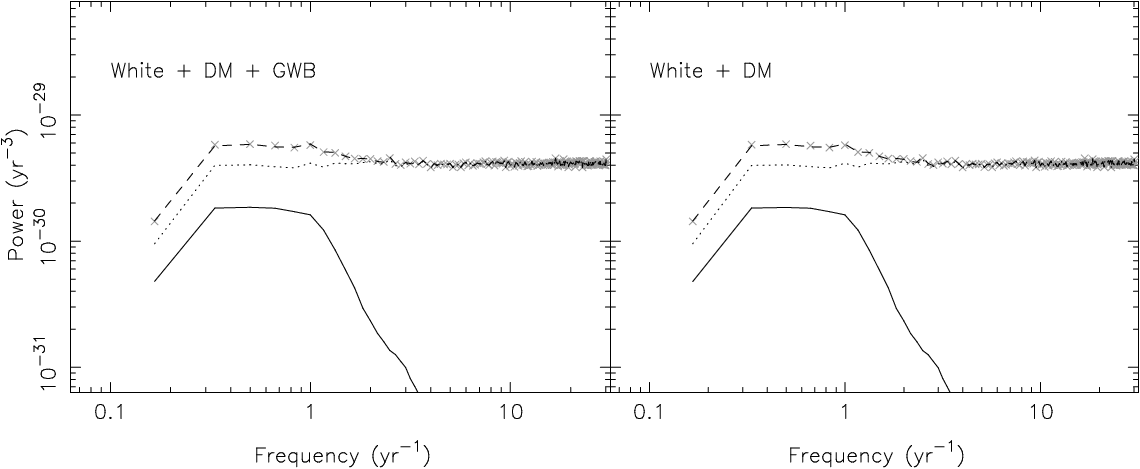}
\caption{
\label{diff_spec}
Average power spectra of the error in DM$(t)$ after fitting to simulations with realistic sampling and uncertainties.
The simulations contained white noise, DM variations and, in the left panel, a model GWB.
The solid black line shows the power spectrum of ${\rm DM}_{\rm est}(t) - {\rm DM}_{\rm in}(t)$.
The dotted line is the power spectrum of the white noise only.
The dashed line is the post-correction power spectrum of the residuals, after subtracting the model GWB signal if present.
The crosses mark the sum of the black line and the dotted line.
}
\end{figure*}

As a direct measure of performance in the estimating DM$(t)$, we compute the difference between the DM estimated from the fit to the residuals, DM$_{\rm est}(t)$, and the DM input in the simulation, DM$_{\rm in}(t)$.
To better compare with the timing residuals, we convert this error in the DM into the error in $t_{\rm DM} (t)$ at $20\,$cm using Equation (\ref{dm_eq}).
Note that, although the residuals were sampled irregularly, the original DM$_{\rm in}$(t) was sampled uniformly on a much finer grid. Furthermore, the estimated DM$_{\rm est}(t)$ is a well defined function that can also be sampled uniformly. Thus it is easy to compute the average power spectrum of this error in $t_{\rm DM} (t)$ as is shown in Figure \ref{diff_spec}. We also plot the spectrum of the initial white noise, and the spectrum of the white noise after correction. If the algorithm is working correctly the white noise after correction should exactly equal the error in $t_{\rm DM} (t)$ plus the white noise before correction, so we have over plotted the sum of these spectra and find that they are identical.

The spectrum of the error in $t_{\rm DM} (t)$ shows the expected behaviour below the corner frequency. Above the corner frequency (where the correction is zero), it falls exactly like the spectrum of $t_{\rm DM} (t)$ itself, i.e., as $f^{-8/3}$. By comparing the right and left panels one can see that the DM correction is independent of the GWB.

\begin{figure}
\includegraphics[width=8cm]{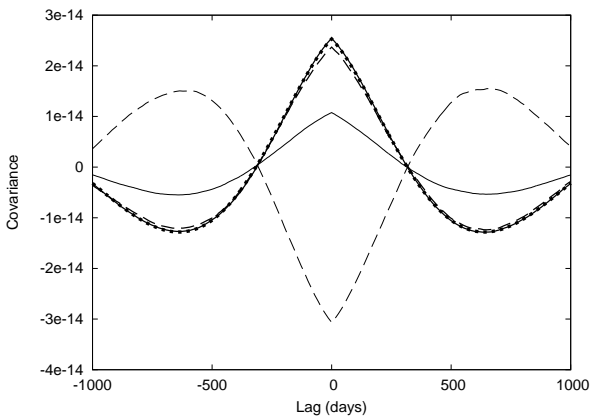}
\caption{
\label{xc_good}
Cross-correlation of post-correction residuals with input model GWB, for the simulations representing PSR J1909$-$3744.
Solid lines show data from the 20-cm wavelength and dashed lines show data from the \b50cm-cm band.
The correction was computed with and without fitting for the common mode, indicated by thick and thin lines respectively.
The auto-correlation of the input GWB is plotted as a dotted line, but it is completely obscured by the heavy solid line for
the cross correlation in the 20-cm band.
}
\end{figure}

We can also demonstrate that the model GWB signal is preserved after DM correction, by cross-correlation of the input model GWB with the post-correction residuals. If the GWB signal is preserved this cross-correlation should equal the auto-correlation of the input GWB signal. We show the auto-correlation of the input and four different cases of the cross-correlation of the output in Figure \ref{xc_good}. The cross-correlations are for two bands (20 and \b50cm$\,$cm shown solid and dashed respectively), and for two different fitting algorithms (with and without $t_{\rm CM} (t)$ shown heavy and light respectively). Again it can be seen that, without fitting for the common-mode $t_{\rm CM} (t)$, a significant portion of the GWB is lost.
In fact, it is apparent from the large negative correlation at \b50cm$\,$cm that the `lost' power is actually transferred from the 20-cm residuals to those at \b50cm$\,$cm.
Although it may be possible to recover this power post-fit, it is not clear how to do this when the GWB and DM signals are unknown.
Finally, we note that when the common mode is used, the \b50cm-cm residuals preserve the GWB just as well as the 20-cm residuals, even though they carry the majority of the DM$(t)$ variation.

\subsection{The Robustness of the Estimator}
The proposed DM correction process is only optimal if the assumptions made in the analysis are satisfied. The primary assumptions are: (1) that there is an unmodelled common-mode signal in the data; (2) that the residuals can be modelled as a set of samples
$t_{oi} (t_j ) = t_{\rm CM} (t_j ) + t_{\rm DM} (t_j ) (\lambda_i / \lambda_{\rm REF} )^2 + t_{wi}(t_j )$; (3) the variances of the samples $t_{wi} (t_j )$ are known.

If Assumption 1 does not hold and we fit for $t_{\rm CM} (t_j )$, then our method would be sub-optimal. However in any pulsar timing  experiment, we must first assume that there is a common-mode signal present. If $t_{\rm CM} (t_j )$ is weak or non-existent then we will have a very low corner frequency and effectively we will not fit for $t_{\rm CM} (t_j )$. So this assumption is tested in every case.

Assumption 2 will fail if there are wavelength dependent terms which do not behave like $\lambda^2$, for example the scattering effect which behaves more like $\lambda^4$. If these terms are present they will corrupt the DM estimate and some scattering effects from longer wavelengths may leak into the shorter wavelengths due to the correction process. However the correction process will not remove any common-mode signal, so signals of interest to PTAs will survive the DM correction unchanged.

Assumption 3 will not always be true a priori. Recent analysis of single pulses from bright MSPs have shown that pulse-to-pulse variations contribute significant white noise in excess of that expected from the formal ToA measurement uncertainty (\citealp{ovh+11,sc12}).
Indeed, many pulsars appear to show some form of additional white noise which is currently unexplained but could be caused by the pulsar, the interstellar medium, or the observing system (see e.g. \citealp{cd85,hbo05}).
In any case, we cannot safely assume that the uncertainties of the timing residuals $\sigma_{ij}$ accurately reflect the white noise level.
If the $\sigma_{ij}$ are incorrect, our fit parameters, $\tilde{t}_c (t)$ and $\tilde{t}_d (t)$, will no longer be minimum variance estimators; however, they will remain unbiased. This means that DM estimation will be unbiased and the DM correction will not remove any GWB (or other common-mode signal) although the correction process may add more white noise than optimal.
It should be noted that if all the $\sigma_{ij}$ were changed by the same factor our DM correction would be unchanged. Fortunately the actual white noise is relatively easy to estimate from the observations because there are more degrees of freedom in the white noise than in the red noise, so in practise we use the estimated white noise rather than the formal measurement uncertainties $\sigma_{ij}$. 

\section{Application to PPTA Observations}
\begin{table}
\caption{\label{ppta_table}Scintillation and dispersion properties for the 20 PPTA pulsars, at a reference wavelength of $20\,$cm.
The scintillation bandwidth ($\nu_0$) and time scale ($\tau_{\rm diff}$) are averaged over a large number of PPTA observations except for values in parenthesis which are are taken from \citeN{yhc+07}. $D_{1000}$ is the value of the structure function at 1000 days and $T_{\rm s}$ is the optimal sampling interval for $t_{\rm DM}(t)$.}
\begin{tabular}{llllc}
\hline \hline
Source   & $\nu_0$ & $\tau_{\rm diff}$ & $D_{1000}$ &$T_{\rm s}$\\
	 & (MHz)   & (s)   & ($\mu s^2$) & (yr) \\
\hline	    
J0437$-$4715 &    1000 &    2486 &     1.6 & 0.2\\
J0613$-$0200 &    1.64 &    4500 &     0.3 & 1\\
J0711$-$6830 &      36 &    1962 &     1.9 & 2\\
J1022+1001 &      65 &    2334 &    0.14 & 2\\
J1024$-$0719 &     268 &    4180 &     6.2 & 1\\
J1045$-$4509 & (0.094) &   (119) &     690 & 0.25\\
J1600$-$3053 &    0.09 &     271 &      24 & 0.5\\
J1603$-$7202 &       5 &     582 &     5.5 & 1\\
J1643$-$1224 &   0.022 &     582 &      65 & 0.5\\
J1713+0747 &      24 &    2855 &    0.31 & 1\\
J1730$-$2304 &    12.4 &    1615 &      20 & 1\\
J1732$-$5049 &     5.4 &    1200 &    10.0 & 1\\
J1744$-$1134 &      60 &    2070 &     1.3 & 1\\
J1824$-$2452A & (0.025) &    (75) &     250 & 0.33\\
J1857+0943 &     5.5 &    1464 &     0.9 & 2\\
J1909$-$3744 &      37 &    2258 &     3.5 & 0.33\\
J1939+2134 &     1.2 &     327 &     8.9 & 0.33\\
J2124$-$3358 &  (1170) & (10705) &     0.4 & 2\\
J2129$-$5721 &    17.1 &    3060 &    0.49 & 2\\
J2145$-$0750 &     195 &    3397 &    0.15 & 2\\

\hline
\end{tabular}
\end{table}

We have applied the new DM correction technique to the PPTA data set \cite{mhb+12}.
Observations of the PPTA are made in three wavelength bands: `$10\,$cm' ($\sim 3100$ MHz); `$20\,$cm' ($\sim 1400$ MHz); and `\b50cm$\,$cm' ($\sim 700$ MHz).
The 10-cm and 20-cm bands have been constant over the entire time span, however the long wavelength receiver was switched from a centre frequency of 685 MHz to 732 MHz around MJD $55030$ to avoid RFI associated with digital TV transmissions.
To allow for changes in the intrinsic pulse profile between these different wavelength bands, we fit for two arbitrary delays between one wavelength band and each of the other bands. However we did not allow an arbitrary delay between 685 and 732 MHz because the pulse shape does not change significantly in that range.

\begin{figure*}
\centerline{\includegraphics[width=19cm,angle=0]{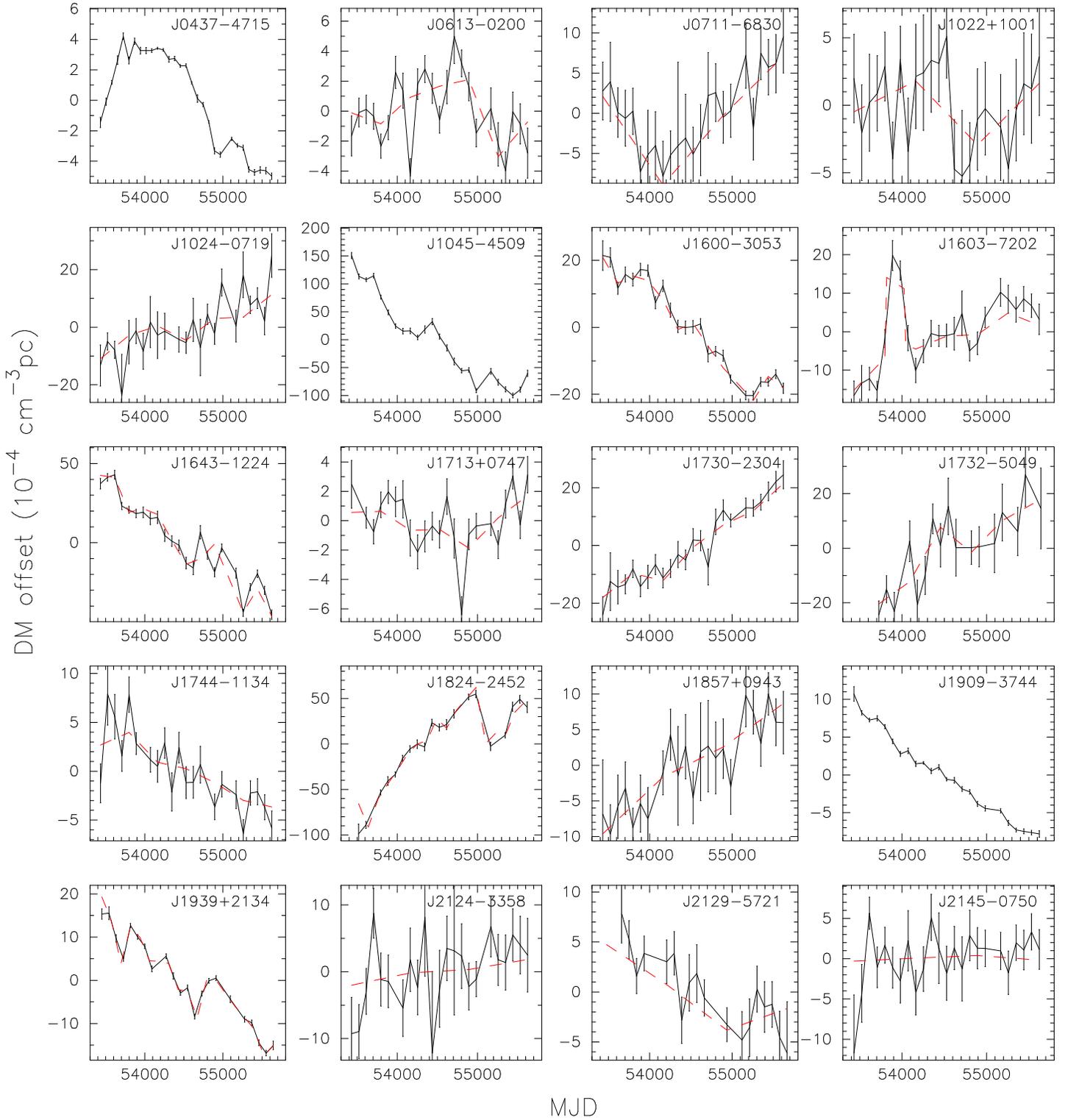}}
\caption{\label{ppta_dm}DM as a function of time for 20 PPTA pulsars. Solid lines show values measured with intervals of 0.25 yr. In the cases where the optimal $T_s$ at a wavelength of $20\,$cm is longer than 0.25 yr, a dashed line is added showing DM$(t_i)$ measured with this time step.}
\end{figure*}

We began our analysis by using the procedure described in Section \ref{technique} to compute pilot estimations of DM$(t)$ and $t_{\rm CM}(t)$ for each of the 20 pulsars, using a sampling interval $T_{\rm s} = 0.25$ yr.
Figure \ref{ppta_dm} shows the DM$(t)$ derived from the above.
Our results are consistent with the measurements made by \citeN{yhc+07} for the $\sim 500$ days of overlapping data, which is expected since they are derived from the same observations.

\subsection{Determining the sampling interval}
\label{sec:corner}
As discussed in Section \ref{technique}, we can use the diffractive time scale $\tau_{\rm diff}$ to predict the magnitude of the DM variations in a given pulsar.
This value can be computed directly from observations, however it is always quite variable on a day to day time scale (see Section \ref{non-dm} for discussion), and for a few pulsars $\tau_{\rm diff}$ approaches the duration of the observations, so it can be hard to measure.
Nevertheless we have obtained an estimate of the average $\tau_{\rm diff}$ from the dynamic spectra for each pulsar, and this is given in Table \ref{ppta_table}. We have not provided an error estimate on the average $\tau_{\rm diff}$ because the variation usually exceeds a factor of two so the values tabulated are very rough estimates.

\begin{figure*}
\centerline{\includegraphics[width=19cm,angle=0]{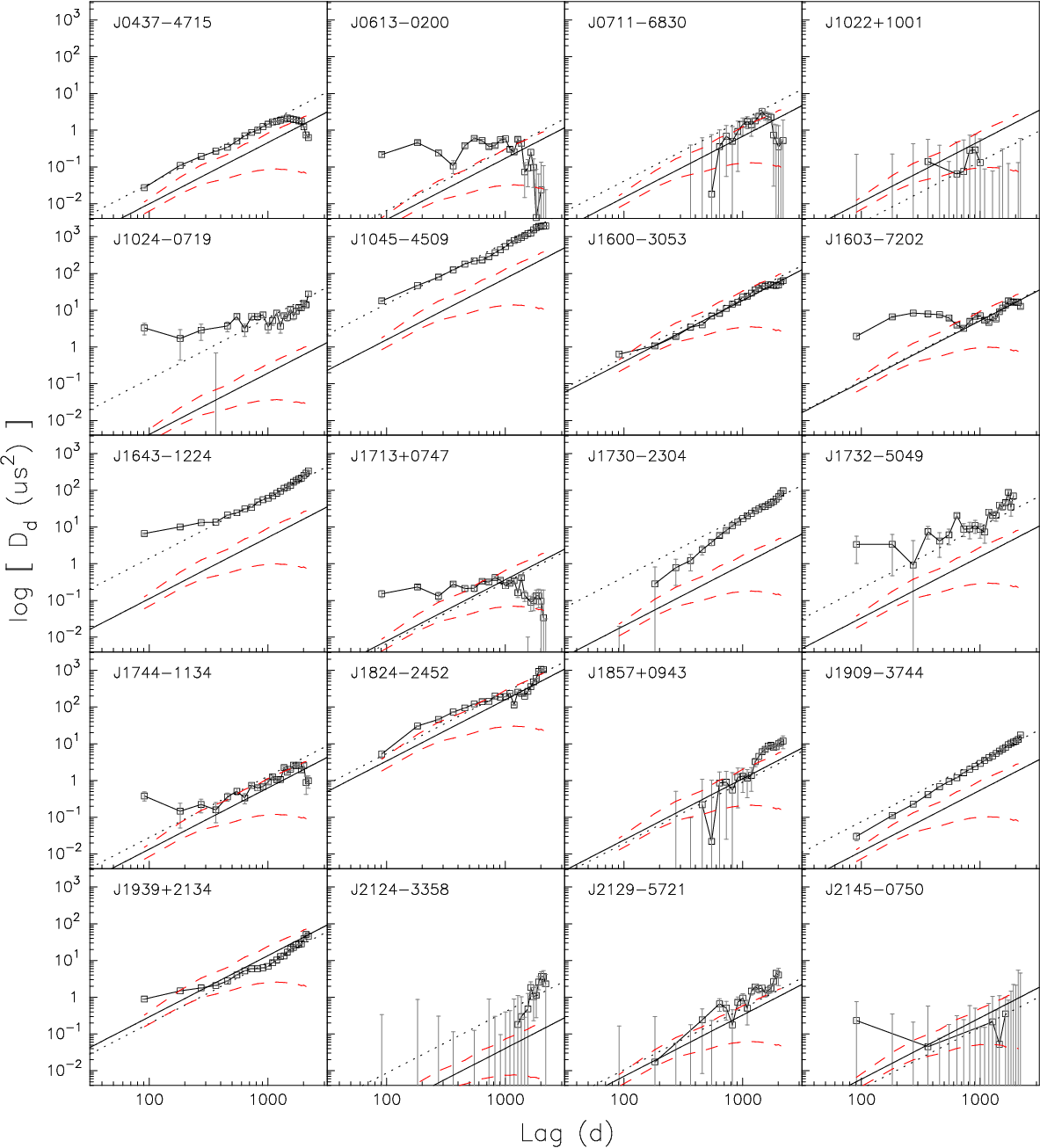}}
\caption{\label{ppta_sf}Structure functions of dispersive delay at $20\,$cm. The square markers indicate the structure function as measured directly from the DM time-series in Figure \ref{ppta_dm}, error bars are derived by simulation of white noise.
The solid lines show the extrapolation from the scintillation time scale $\tau_{\rm diff}$ assuming a Kolmogorov spectrum, dashed lines mark the region occupied by 68\% of simulated data sets having Kolmogorov noise with the same amplitude. These lines indicate the uncertanty in the measured structure functions resulting from the finite length of the data sets.
The dotted lines show a Kolmogorov spectrum with the amplitude set to match the real data at a lag of 1000 days.
}
\end{figure*}

We also computed the structure function $D_{\rm TDM}$ directly from the $t_{\rm DM}$ values.
These structure functions, scaled to delay in $\mu$s$^2$ at $20\,$cm, and those estimated from $\tau_{\rm diff}$, are shown in Figure \ref{ppta_sf}.
The value $D_{\rm TDM}(1000\,{\rm days})$ is given in Table \ref{ppta_table}.

For each pulsar we also make an estimate of the white noise power directly from the power spectrum of the residuals.
The estimates of $P_w$ at each wavelength are given in Table \ref{psd_table}.

We then use the $D_{\rm TDM}$ estimates and Equation (\ref{eq:pd}) to generate a model power spectrum $P_{\rm TDM}(f)$ at the reference wavelength ($20\,$cm) for each pulsar. These assume a Kolmogorov spectral exponent. From these model spectra and the corresponding $P_{w,TDM}$, tabulated in Table \ref{psd_table}, we determine the corner frequency and the corresponding sample interval $T_s$ for DM for each pulsar. As we do not have any a priori knowledge of $t_{\rm CM}$ for the PPTA pulsars we choose the same sample interval for $t_{\rm CM}$ as for $t_{\rm DM}$.

\subsection{Results}
\begin{table}
\caption{\label{f2table}Impact of the DM corrections on the timing parameters, as determined in the 20-cm band. For each pulsar we present the change in $\nu$ and $\dot\nu$ due to the DM correction, relative to the measurement uncertainty, and the ratio of the RMS of the residuals before ($\Sigma_{\rm pre}$) and after ($\Sigma_{\rm post}$) DM correction.
Also included is the ratio of the power spectral density before ($\bar{P}_{\rm pre}$) and after ($\bar{P}_{\rm post}$) correction, averaged below $f_c$.
The final column indicates if we believe that the DM corrections have `improved' the data set for the purpose of detecting common-mode red signals.
}
\begin{center}
\begin{tabular}{lcccr@{-}lc}
\hline
PSR & $ \frac{|\Delta \nu|}{\sigma_\nu}$ & $ \frac{|\Delta \dot\nu|}{\sigma_{\dot\nu}}$ & $\frac{\Sigma_{\rm post}}{\Sigma_{\rm pre}}$ & \multicolumn{2}{c}{$\frac{\bar{P}_{\rm post}}{\bar{P}_{\rm pre}}$}  &Imp. \\
\hline
\hline
J0437$-$4715 &   92 &   48 & 0.6 &0.15&0.25 & Y \\
J0613$-$0200 & 0.16 &  2.9 & 1.1 & 0.3&1.2 & y \\
J0711$-$6830 &  3.9 &  5.5 & 1.0 & 0.4&1.6 & y \\
J1022+1001 &  1.4 &  0.3 & 1.0 & 0.6&2.6 & n \\
J1024$-$0719 &    1 & 0.91 & 1.0 & 0.2&0.7 & Y \\
J1045$-$4509 &   28 &   11 & 0.7 &0.22&0.39 & Y \\
J1600$-$3053 &   35 & 0.51 & 1.0 & 0.4&0.8 & Y \\
J1603$-$7202 &  2.4 &  2.5 & 1.0 & 0.2&0.9 & Y \\
J1643$-$1224 &   11 & 0.73 & 1.7 & 1.3&3.1 & N \\
J1713+0747 &  3.2 &  6.2 & 1.0 & 0.2&0.7 & Y \\
J1730$-$2304 &  6.5 &  1.8 & 1.1 & 0.9&3.2 & n \\
J1732$-$5049 &  2.6 &  2.8 & 1.0 & 0.4&1.4 & y \\
J1744$-$1134 &  5.4 & 0.48 & 1.0 & 0.5&2.0 & n \\
J1824$-$2452A &   24 &   31 & 0.7 &0.29&0.56 & Y \\
J1857+0943 &  4.3 &    1 & 1.0 & 0.2&1.0 & y \\
J1909$-$3744 &   28 &    5 & 1.0 &0.44&0.79 & Y \\
J1939+2134 &   13 &  1.7 & 0.7 &0.34&0.67 & Y \\
J2124$-$3358 & 0.25 & 0.056 & 1.0 & 0.5&1.9 & y \\
J2129$-$5721 &    3 &  2.1 & 1.1 & 0.7&2.8 & n \\
J2145$-$0750 & 0.22 & 0.18 & 1.0 & 0.2&1.0 & y \\

\hline
\end{tabular}
\end{center}
\end{table}

The  measured DM$(t)$ sampled at the optimal interval $T_s$ is overlaid on the plot of the pilot analysis with $T_s$ = 0.25 yr on Figure \ref{ppta_dm}. It is not clear that there are measurable variations in DM in PSRs J1022+1001, J2124$-$3358 or J2145$-$0750, but one can see that there are statistically significant changes with time for the other pulsars.  In general, the `optimally sampled' time series (dashed line) follows the DM trend with less scatter.
However, there are some significant DM fluctuations that are not well modelled by the smoother time-series.
In particular we do not model the significant annual variations observed in PSR J0613$-$0200, and we must add a step change to account for the 250 day increase in DM observed in PSR J1603$-$7202 (these features are discussed more fully in Section \ref{theism}).
These variations do not follow the Kolmogorov model that was used to derive the optimal sampling rate, and therefore we must use a shorter $T_s$ so we can track these rapid variations.
These results illustrate the importance of making a pilot analysis before deciding on the sample interval. The ISM is an inhomogeneous turbulent process and an individual realisation may not behave much like the statistical average. The DM$(t)$ for PSR J1909$-$3744 is also instructive. It is remarkably linear over the entire observation interval. This linearity would not be reflected in the timing residuals at a single wavelength because a quadratic polynomial is removed in fitting the timing model. It can only be seen by comparing the residuals at different wavelengths. Such linear behaviour implies a quadratic structure function and a power spectrum steeper than Kolmogorov.

\subsection{Performance of DM Correction}

The simplest and most widely used metric for the quality of timing residuals is the RMS of the residuals. Thus a natural
measure of the performance of DM correction would be the ratio of the RMS of the 20-cm residuals before and after DM correction. 
This ratio is provided in Table \ref{f2table}.
However, for most of these pulsars, the RMS is dominated by the white noise and so does not change appreciably after DM correction. Furthermore much of the effect of DM$(t)$ variations is absorbed by fitting for the pulse frequency and its derivative.
Thus the ratio of the RMS before and after DM correction is not a very sensitive performance measure. 
As noted by \citeN{yhc+07}, the DM correction has a significant effect on the pulsar spin parameters, which can give an indication of the magnitude of the DM correction. Table \ref{f2table} lists the change in $\nu$ and $\dot\nu$, as a factor of the measurement uncertainty, caused by applying the DM correction.
However, there are systematic uncertainties in the estimation of the intrinsic values of $\nu$ and $\dot\nu$ that may be greater than the error induced by DM variations.

Judging the significance of the DM corrections depends on the intended use of the data set.
Since a major goal of the PPTA is to search for common-mode red signals, we choose to consider the impact of the DM corrections on the low frequency noise.
In principal, the DM correction should reduce the noise at frequencies below $f_c$, and therefore we have estimated the ratio of the pre- and post-correction power spectrum of the 20-cm residuals, averaged over all frequencies below $f_c$.
We caution that the spectral estimates are highly uncertain, and for many pulsars we average very few spectral channels so the error is non-Gaussian.
Therefore, we present these ratios in Table \ref{f2table} as an estimated $68\%$ uncertainty range, determined assuming the spectral estimates are $\chi^2$-distributed with mean and variance equal to the measured mean power spectral density.

There are 9 pulsars for which the DM correction appears to significantly reduce the low frequency noise, and therefore increases the signal-to-noise ratio for any common-mode signal in the data.
These pulsars are listed with a `Y' in Table \ref{f2table}.
There are 10 pulsars for which the change in low frequency power is smaller than the uncertainty in the spectral estimation and so it is not clear if the DM correction should be performed.
Table \ref{f2table} indicates these pulsars with a `y' or `n', with the former indicating that we believe that the DM correction is likely to improve the residuals.
However, the DM correction fails to `improve' PSR J1643$-$1224 under any metric, even though we measure considerable DM variations (see Figure \ref{ppta_dm}).
As discussed in Section \ref{non-dm}, we believe that this is due to variations in scattering delay entering the DM correction and adding considerable excess noise to the corrected residuals.

\section{Scattering and DM Correction}

\label{non-dm}
\label{longer_wavelength}
\begin{figure}
\begin{center}
\includegraphics[width=9.0cm]{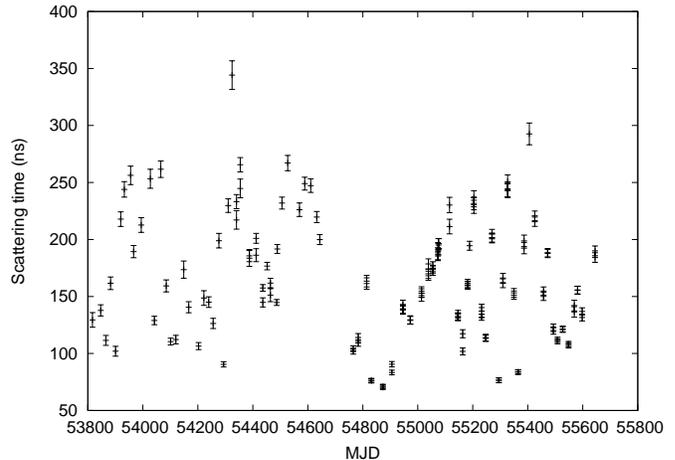}
\caption{
\label{scatt_delay}
Diffractive scattering delay, $t_0$, measured from scintillation bandwidth, $\nu_0$, in observations of PSR J1939+2134 at a wavelength of $20\,$cm.
The error bars are derived from the fit for $\nu_0$ and so are roughly proportional to $t_0$.
}
\end{center}
\end{figure}

The most important effect of the ISM on pulsar timing is the group delay caused by the dispersive plasma along the line of sight.
However small scale fluctuations in the ISM also cause angular scattering by a diffractive process. This scattering  causes a time delay $t_0 \approx 0.5 \theta_0^2 L/c$, where $\theta_0$ is the RMS of the scattering angle and $L$ is the distance to the pulsar. This can be significant, particularly at longer wavelengths, because it varies much faster with $\lambda$ than does the dispersive delay - approximately as $\lambda^{4}$. 
In homogeneous turbulence one would expect this parameter to be relatively constant with time.
If so, the delay can be absorbed into the pulsar profile and it will have little effect on pulsar timing. However if the turbulence is inhomogeneous the scattering delay may vary with time and could become a significant noise source for pulsar timing.
We can study this effect using the PPTA pulsar PSR J1939+2134.
Although this pulsar is unusual in some respects, the scattering is a property of the ISM, not the pulsar, and the ISM in the direction of PSR J1939+2134 can be assumed to be typical of the ISM in general. PSR J1939+2134 is a very strong source and the observing parameters used for the PPTA are well-suited to studying its interstellar scattering. The time delay, $t_0$, can be estimated from the bandwidth of the diffractive scintillations, $\nu_0$, in a dynamic spectrum using the relationship $t_0 = 1/2 \pi\nu_0$.  In fact it is extremely variable, as can be seen in Figure \ref{scatt_delay}.
The RMS of $t_0$ (52 ns at $20\,$cm) is about 28\% of the mean.
We can expect this to increase by a factor of $(1400\,{\rm MHz}/700\,{\rm MHz})^{4} = 16$ at \b50cm$\,$cm. Thus in the \b50cm-cm ToAs there will be delays with RMS variations of $\sim830\,$ns, which do not fit the dispersive $\lambda^2$ behaviour.
This will appear in the estimate of $t_{\rm DM}$ at $20\,$cm, attenuated by a factor of $((1400\,{\rm MHz}/700\,{\rm MHz})^2 -1) = 3$ (Equation \ref{eq1}).
Therefore the DM correction will bring scattering noise from the 50-cm band to the 20-cm band with RMS variation $\sim 270\,$ns, 5.3 times larger than the scattering noise intrinsic to the 20-cm observations.
This analysis is corroborated by the structure function of DM for this pulsar shown in Figure \ref{ppta_sf}, which shows a flattening to about 1$\,\mu$s$^2$ at small time lags.
This implies a white process with RMS variations of about $500\,$ns, consistent with that expected from scattering.

We have correlated the variations in $t_0$ with the 20-cm residuals before correction and find 18\% positive correlation.
This is consistent with the presence of $52\,$ns of completely correlated noise due to $t_0$ added to the ToA measurement uncertainty of the order of $200\,$ns. PSR J1939+2134 is known to show ToA variations that are correlated with the intensity scintillations \cite{cbl+95} but are much stronger than expected for  homogeneous turbulence \cite{crg+10}. Thus we are confident that the observed variation in $t_0$ is showing up in the 20-cm residuals. We expect that contribution to increase in the DM corrected residuals to about 300 ns. However this is very difficult to measure directly because the DM correction is smoothed and the \b50cm-cm observations are not simultaneous with the 20-cm observations.

This effect increases rapidly with longer wavelength.
If we had used 80-cm observations for DM correction, the RMS at 80~cm would have been $\sim10\,\mu$s and this would have been reduced by a factor of 12 to an RMS of $800\,$ns in the corrected $20\,$cm residuals.
Clearly use of low frequency antennas such as GMRT \cite{jr06} or LOFAR \cite{sha+11} for correcting DM fluctuations in PTAs will have to be limited to weakly scattered pulsars.
This is an important consideration, but it should be noted that the four PPTA pulsars that provide the best timing are all scattered much less than J1939+2134 - all could be DM corrected with 80-cm observations or even with longer wavelengths.
On the other hand there are four PPTA pulsars that are scattered 20 to 80 times more strongly than J1939+2134 and even correction with \b50cm-cm data causes serious increases in the white noise.

An extreme example is PSR J1643$-$1224. Under the above assumption, the expected white noise (2.0~$\mu$s) due to scattering at $20\,$cm exceeds the radiometer noise (0.63 $\mu$s). The white scattering noise at \b50cm$\,$cm is much larger (32 $\mu$s) and about a third of this makes its way into the DM-corrected residuals at $20\,$cm.
This is also corroborated by the structure function for this pulsar in Figure \ref{ppta_sf}, which shows a flattening to about $10\,\mu$s$^2$ at small lags. This implies a white process with RMS variation of $\sim3\,\mu$s which is consistent with that expected from scattering.
Indeed, this pulsar is the only pulsar with significant DM variations for which the DM correction increases the noise in the 20-cm residuals under all metrics.
It is also important to note that observing this source at the same frequencies with a more sensitive telescope will not improve the signal to noise ratio, because the noise, both before and after DM-correction, is dominated by scattering. However using a more sensitive telescope could improve matters by putting more weight on observations at $10\,$cm, where scattering is negligible.

Finally however, we note that the usefulness of long wavelength observations would be greatly improved if one could measure and correct for the variation in scattering delays.
This may be possible using a technique such as cyclic spectroscopy, however this has only been done in ideal circumstances and with signal-to-noise ratio such that individual pulses are detectable \cite{dem11}.
It is still unclear if such techniques can be generalised to other observations, or if this can be used to accurately determine the unscattered ToA.

\section{Scheduling for DM Correction}

If there were no DM variation, one would spend all the observing time at the wavelength for which the pulsar has the greatest ToA precision (see \citealp{mhb+12} for discussion on choice of observing wavelength).
The reality is, of course, that we need to spend some of the time observing at other wavelengths to correct for DM variations.
In this section we will present a strategy for choosing the observing time at each wavelength, attempting to optimise the signal to noise ratio of the common-mode signal, $t_{\rm CM}$.
We take the PPTA observations at Parkes as our example, but this work can easily be generalised to any telescope.

At Parkes it is possible to observe at wavelengths of 10 and \b50cm$\,$cm simultaneously because the ratio of the wavelengths is so high that the shorter wavelength feed can be located co-axially inside the longer wavelength feed.
However the 20-cm receiver does not overlap with either 10 or \b50cm$\,$cm and so must be operated separately.

As noted earlier we can write $\tilde{t}_{\rm CM} = \sum_i b_i t_{oi}$ so the variance of $t_{\rm CM}$ is given by 
$\sigma_{\rm TCM}^2 = \sum_i b_i^2 \sigma_i^2$.
However, the solution is not linear in the $\sigma_i^2$ terms because they also 
appear in the $b_i$ coefficients. 
At present, observing time at Parkes is roughly equal between the two receivers, so we use the existing power spectral densities as the reference.
We will assume that the total 
observing time is unity and the time devoted to 10 and \b50cm$\,$cm,
which are observed simultaneously, is $0 \leq \alpha \leq 1$.

The variance of any $t_{oi}$ is inversely proportional to the corresponding observing 
time. We use $t_{o,20}$ as the reference because it usually has the smallest TOA uncertainty in PPTA observations.
Therefore, we define $\sigma_{20}^2 = 1/(1 - \alpha)$ as the reference. Then we assume that, with 
equal observing time $\sigma_{10} = x \sigma_{20}$ and $\sigma_{50} = y \sigma_{20}$, so as
scheduled we would have $\sigma_{10} = x/\alpha$ and $\sigma_{50} = y/\alpha$.
We can then determine the 
increase in white noise caused by correcting for dispersion as a function of $\alpha$, the time 
devoted to 10 and \b50cm$\,$cm. The results are shown for all the PPTA pulsars in Table \ref{psd_table}.
One can see that all the pulsars are different and the optimal strategies range from $\alpha \approx 0.2$
to $\alpha \approx 1.0$ (i.e. 100\% of time spent using the dual 10 and \b50cm$\,$cm system).
For the four `best' pulsars, PSRs J0437$-$4715, J1713+0747, J1744$-$1134, and J1909$-$3744, the optimal strategy has $\alpha > 0.7$.

This suggests that a useful improvement to PTA performance could come from deploying broadband receivers, so that correction for DM$(t)$ can be done with a single observation.
This also has the benefit of reducing the difficulties of aligning pulsar profiles measured with different receivers at different times, and would therefore allow for more accurate measurement of DM variations.

\section{The Interstellar Medium}
\label{theism}

The PPTA DM$(t)$ observations provide an interesting picture of the ionised ISM on au spatial scales. The overall picture can be seen in Figures \ref{ppta_dm} and \ref{ppta_sf}. In Figure \ref{ppta_dm} it is apparent that 17 of the 20 PPTA pulsars have measurable DM$(t)$ variations. In Figure \ref{ppta_sf} it can be seen that 13 of these 17 show power-law structure functions, as expected in the ensemble average
Of these, eight are roughly consistent with an extrapolation from the diffractive scales at the Kolmogorov spectral exponent, an average dynamic range of 4.8 decades. However five are considerably higher than is predicted by a Kolmogorov extrapolation.
They may be locally Kolmogorov, i.e. an inner scale may occur somewhere between the diffractive scale and the directly measured scales of 100 to 2000 days, but establishing this would require a detailed analysis of the apparent scintillation velocity which is beyond the scope of this paper. Two of these five pulsars, J1045$-$4509 and J1909$-$3744, were already known to be inconsistent with a Kolmogorov spectral exponent \cite{yhc+07}, and it is clear, with the additional data that are now available, that J1024$-$0719, J1643$-$1224 and J1730$-$2304 should be added to this list. When the spatial power spectrum of a stochastic process is steeper than the Kolmogorov power-law, it can be expected to be dominated by linear gradients and show an almost quadratic structure function. Indeed inspection of Figure \ref{ppta_dm} shows that the 5 steep spectrum pulsars all show a strong linear gradient in DM$(t)$.


The time series DM$(t)$ shown in Figure  \ref{ppta_dm} often show behaviour that does not look like a homogeneous stochastic process. For example, PSR J1603$-$7202 shows a large increase for $\sim$250 days around MJD 54000 and J0613$-$0200 shows clear annual modulation. The increase in DM for J1603$-$7202 suggests that a blob of plasma moved through the line of sight. If we assume the blob is halfway between the pulsar and the Earth, the line of sight would have moved by about 0.5 au in this time, and if the blob were spherical it would need a density of $\sim$200 cm$^{-3}$. This value is high, but comparable to other density estimates for au-scale structure based on `extreme scattering events'  \cite{fdjh87,cbl+93}.

We computed the power spectra of DM$(t)$ for all the pulsars to see if the annual modulation that is clear by eye in PSR J0613$-$0200 is present in any of the other pulsars. 
For four pulsars we find a significant ($>5$-$\sigma$) detection of an annual periodicity, PSRs J0613$-$0200, J1045$-$4509, J1643$-$1224 and J1939+2134.

\begin{figure}
\begin{center}
\includegraphics[width=6.5cm]{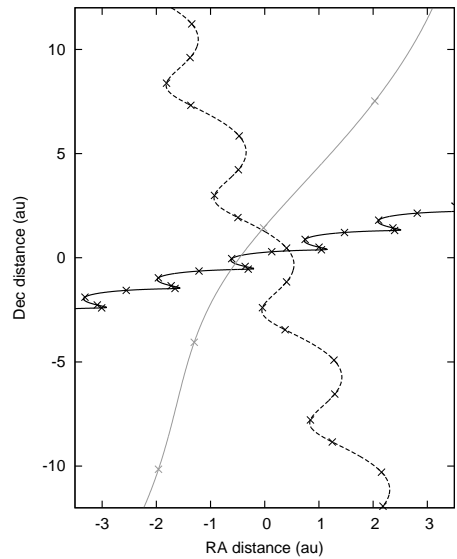}
\caption{
\label{trajectory}
Trajectories through the ISM of the line of sight to PSRs J0613$-$0200 (dashed line), J1643$-$1224 (solid black line) and J1909$-$3744 (grey line). It was assumed that the scattering takes place half way between the pulsar and the Earth and the motion of the plasma was neglected. The trajectories are marked with a cross at the DM sampling interval of 0.25 yr.
}
\end{center}
\end{figure}

The most likely explanation for the annual variation in DM$(t)$ is the annual shift in the line of sight to the pulsar resulting from the orbital motion of the Earth.
The trajectory of the line of sight to three example PPTA pulsars are shown in Figure \ref{trajectory}.
The relatively low proper motion and large parallax of the PPTA pulsars means that the trajectory of the line of sight to many of the PPTA pulsars show pronounced ripples.
However, unless the trajectory is a tight spiral, the annual modulation will only be significant if there is a persistent gradient in the diffractive phase screen.

The presence of persistent phase gradients and annual modulation in J1045$-$4509 and J1643$-$1224 is not surprising because the ISM associated with each of these pulsars has a steeper than Kolmogorov power spectrum.
Indeed, the measured DM$(t)$ for these pulsars do show a very linear trend, which in itself evidence for a persistent phase gradient.
The other steep spectrum pulsars, J1024$-$0719, J1730$-$2304 and J1909$-$3744, have higher proper motion, which reduces the amplitude of the annual modulation relative to the long term trend in DM$(t)$. We note that the spectral analyses for PSRs J1024$-$0719 and J1909$-$3744 suggest annual periodicities, and it may be possible to make a significant detection by combining the PPTA data with other data sets.

PSR J1939+2134 does not show a steep spectrum, however its proper motion is very low compared to its parallax, and therefore the trajectory spirals through the ISM, reducing the requirement for a smooth phase screen.
The annual modulation of J0613$-$0200 may be somewhat different, since it does not have a steep spectrum and although the proper motion is small the trajectory does not spiral (see Figure \ref{trajectory}).
This suggests that for J0613$-$0200 the turbulence could be anisotropic with the slope of the gradient aligned with the direction of the proper motion. Anisotropic structures are believed to be quite common in the ISM~\cite{crsc06,bmg+10}. However one can imagine various other ways in which this could occur, particularly in an inhomogeneous random process, and inhomogeneous turbulence on an au spatial scale is also believed to be common in the ISM~\cite{smc+01,crsc06,bmg+10}.

Persistent spatial gradients will cause a refractive shift in the apparent position of the pulsar, and because of dispersion the refraction angle will be wavelength dependent.
This refractive shift appears in the timing residuals as an annual sine wave which changes in amplitude like $\lambda^2$. When the DM$(t)$ is corrected this sine wave disappears and the inferred position becomes the same at all wavelengths.
These position shifts are of order $10^{-4}\,(\lambda/20\,{\rm cm})^2$ arcseconds for all four pulsars.

Note that the trajectory of the lines of sight shown on Figure \ref{trajectory} may appear quite non-sinusoidal, but the annual modulation caused by the Earth's orbital motion in a linear phase gradient will be exactly a sine wave superimposed on a linear slope due to proper motion. This will not generate any higher harmonics unless the structure shows significant non-linearity on an au scale. We do not see second harmonics of the annual period, which suggests that the spatial structure must be quite linear on an au scale.

Annual variations in DM are also observed in pulsars for which the line of sight passes close to the Sun because of free electrons in the solar wind \cite{ojs07,xchm12}.
In the PPTA, a simple symmetric model of the solar wind is used to remove this effect, but this is negligible for most pulsars.
For the three pulsars where it is not negligible, the effect of the solar wind persists only for a few days at the time when the line of sight passes closest to the Sun.
Neither the magnitude, phase nor shape of the variations seen in our sample can be explained by an error in the model of the solar wind.
Changes in ionospheric free electron content can be ruled out for similar reasons.

In summary the ISM observations are, roughly speaking, consistent with our present understanding of the ISM. However the data will clearly support a more detailed analysis, including spectral modelling over a time scale range in excess of $10^5$ from the diffractive scale to the duration of the observations. It may also be possible to make a 2-dimensional spatial model of the electron density variations for some of the 20 PPTA pulsars.
Although this would be useful for studying the ISM and in improving the DM correction, such detailed modelling is beyond the scope of this paper.

\section{Conclusions}

We find that it is necessary to approach the problem of estimating and correcting for DM$(t)$ variations iteratively, beginning with a pilot analysis for each pulsar and refining that analysis as the properties of that pulsar and the associated ISM become clearer. Each pulsar is different and the ISM in the line of sight to each pulsar is different. The optimal analysis must be tailored to the conditions appropriate for each pulsar and according to the application under consideration.

We sample the DM$(t)$ just often enough that the variations in DM are captured with the minimum amount of additional white noise. Likewise, we must also sample a common-mode signal $t_{CM}(t)$ at the appropriate rate. In this way we can correct for the DM variations at frequencies where it is necessary, and we can include $t_{CM}(t)$ at frequencies where it is necessary, but not fit for either 
at frequencies where the signal is dominated by white noise.

By including the common-mode signal in the analysis we preserve the wavelength-independent signals of interest for pulsar timing arrays and we improve the estimate of the pulsar period and period derivative. Without estimating the common mode, a significant fraction of wavelength-independent signals, such as: errors in the terrestrial clocks; errors in the planetary ephemeris; and the effects of gravitational waves from cosmic sources, would have been absorbed into the DM correction and lost.

We have applied this technique to the PPTA data set, which improves its sensitivity for the detection of low frequency signals significantly. The estimated DM$(t)$ also provides an unparallelled measure of the au scale structure of the interstellar plasma. In particular it confirms earlier suggestions that the fluctuations often have a steeper than Kolmogorov spectrum, which implies that an improved physical understanding of the turbulence will be necessary. We also find that persistent phase gradients over au scales are relatively common and are large enough to cause significant errors in the apparent positions of pulsars unless DM corrections are applied.

\section*{Acknowledgements} 

This work has been carried out as part of the Parkes Pulsar Timing Array project. GH is the recipient of an Australian Research Council QEII Fellowship (project DP0878388), The PPTA project was initiated with support from RNM's Federation Fellowship (FF0348478). The Parkes radio telescope is part of the Australia Telescope which is funded by the Commonwealth of Australia for operation as a National Facility managed by CSIRO.

\bibliographystyle{mnras}
\bibliography{journals,myrefs,modrefs,psrrefs,crossrefs} 

\appendix

\section{Constrained least squares fitting in Tempo2}
\label{appendix2}
The least squares problem of
fitting the timing model to the residuals can be written in matrix form as
\begin{equation}
\bf R = {\bf M P + E}.
\end{equation}
Here $\bf R$ is a column vector of the timing residuals, $\bf P$ is a column vector of fit parameters, including DM$(t_j)$ and $t_{\rm CM}(t_j)$ as well as the other timing model parameters.
$\bf M$ is a matrix describing the timing model and $\bf E$ is a column vector of errors.
The least-squares algorithm solves for $\bf P$, matching ${\bf M P}$ to $\bf R$ with a typical accuracy of $\bf E$.

The sampled time series DM$(t_j)$ and $t_{\rm CM}(t_j)$ are covariant with the timing model, so they must be constrained to
eliminate that covariance or the least squares solution will fail to converge on a unique solution.
These constraints have the form of linear equations of DM$(t_j)$ and $t_{\rm CM}(t_j)$, such as:
$\sum DM(t_j) = 0$;  $\sum t_{\rm CM}(t_j) = 0$; $\sum t_j t_{\rm CM}(t_j) = 0$, $\sum t_j^2 t_{\rm CM}(t_j) = 0$;
$\sum \sin(\omega t_j) t_{\rm CM}(t_j) = 0$; $\sum \cos(\omega t_j) t_{\rm CM}(t_j) = 0$; etc. 
Augmented with these equations, the least-squares problem becomes
\[
\begin{bmatrix} \bf R \\ \bf C \end{bmatrix} = \begin{bmatrix} \bf M \\ \bf B \end{bmatrix} \bf P + 
\begin{bmatrix} \bf E \\ \bf \epsilon \end{bmatrix},
\]
where $\bf B$ is a matrix describing the constraints, $\bf \epsilon$ is a column vector of weights for the constraints.
In our case $\bf C = 0$, though it need not be in general.
The least-squares solution will then find a vector $\bf P$ that matches both ${\bf M P}$ to $\bf R$, with a typical accuracy of $\bf E$, and also matches ${\bf B P}$ to $\bf C$, with a typical accuracy of $\bf \epsilon$. By making $\bf \epsilon$ very small we can enforce the constraints with high accuracy. This scheme has been called `the method of weights'~\cite{gou96}.

If the uncertainties in the estimates of DM$(t_j)$ and $t_{\rm CM}(t_j)$ are not expected to be equal, for instance if the different observing wavelengths are irregularly sampled and the ToA uncertainties are variable across sampling windows, then it can be advantageous to use weighted constraints.
Then the constraints take the form $\sum W_j DM(t_j) = 0$, and we need to estimate the uncertainties of the parameters to obtain the optimal weights.
These uncertainties can be determined from the least-squares solution in which the timing residuals are described purely by Equation (\ref{eq4}).
This problem is linear and the covariance matrix of the parameters can be written in closed form without even solving for the parameters. The diagonal elements of the covariance matrix are the variances of the parameters and the weights, $W_j$, are the inverse of the square roots of the corresponding variances.

\section{Relation between the structure function and power-spectral density}
\label{appendix1}
The structure function $D(\tau)$, of a time series $y(t)$, is well defined if $y(t)$ has stationary differences
\begin{equation}
D(\tau) = \langle (y(t) - y(t + \tau ))^2 \rangle .
\end{equation}
If $y(t)$ is wide-sense stationary $D(\tau)$ can be written in terms of the auto covariance $C(\tau)$ by expansion of the square
\begin{eqnarray}
D(\tau) &=& \langle y(t)^2 \rangle + \langle y(t+\tau)^2 \rangle - 2\langle y(t)y(t+\tau) \rangle \nonumber \\
 &=& 2 (C(0) - C(\tau))
\end{eqnarray}
If $y(t)$ is real valued then by the Wiener-Khinchin theorem,
\begin{equation}
C(\tau)=  \int_{0}^{\infty} \! \cos(2\pi f\tau)\, P(f)\,{\rm d}f,
\end{equation}
where $P(f)$ is the one-sided power spectral density of $y(t)$. Thus we can then write the structure function in terms of the 
power spectral density as
\begin{equation}
D(\tau) = A \int_{0}^{\infty} \! 2(1-\cos(2\pi f\tau)) P( f)\,{\rm d}f,
\end{equation}
It should be noted that this expression for $D(\tau)$ is valid if $D(\tau)$ exists. It is not necessary that $C(\tau)$ exist.
For the case of a power-law, $P(f)=A f^{-\alpha}$, we can change variables using $x=f\tau$, and obtain
\begin{equation}
D(\tau) = \tau^{\alpha-1} A \int_{0}^{\infty} \! 2(1-\cos(2\pi x))\, x^{-\alpha}\,{\rm d}x.
\end{equation}
The integral (Int) above converges if $1<\alpha<3$, yielding
\begin{equation}
{\rm Int.} = 2^\alpha \pi^{\alpha-1} \sin(-\alpha\pi/2) \Gamma(1-\alpha), 
\end{equation}
where $\Gamma$ is the Gamma function.
Thus for Kolmogorov turbulence, with exponent $\alpha=8/3$, we have ${\rm Int} \simeq 89.344$ and the power spectrum can be written
\begin{equation}
\label{eq:sf2psd}
P(f) \simeq 0.0112 D(\tau)\tau^{-5/3} f^{-8/3}.
\end{equation}

\end{document}